\documentclass[12pt]{article}
\usepackage{graphicx}

\newcommand\pubnumber{NuPhys2015-Alvarez-Ruso}
\newcommand\pubdate{May 16, 2016}

\def\Title#1{\begin{center} {\Large #1 } \end{center}}
\def\Author#1{\begin{center}{ \sc #1} \end{center}}
\def\Address#1{\begin{center}{ \it #1} \end{center}}

\newcommand\pubblock{\rightline{\begin{tabular}{l} \pubnumber\\
         \pubdate  \end{tabular}}}
\newenvironment{Abstract}{\begin{quotation}  }{\end{quotation}}
\newenvironment{Presented}{\begin{quotation} \begin{center} 
             PRESENTED AT\end{center}\bigskip 
      \begin{center}\begin{large}}{\end{large}\end{center} \end{quotation}}
\def\Acknowledgements{\bigskip  \bigskip \begin{center} \begin{large}
             \bf ACKNOWLEDGEMENTS \end{large}\end{center}}

\def\beq{\begin{equation}}
\def\eeq#1{\label{#1}\end{equation}}
\def\eeqn{\end{equation}}

\def\beqa{\begin{eqnarray}}
\def\eeqa#1{\label{#1}\end{eqnarray}}
\def\eeqan{\end{eqnarray}}

\let\bar=\overbar

\def\Dslash{\not{\hbox{\kern-4pt $D$}}}
\def\dslash{\not{\hbox{\kern-2pt $\del$}}}

\def\ee{e^+e^-}

\def\msb{{\bar{\ssstyle M \kern -1pt S}}}

\def\be{\begin{equation}}
\def\ee{\end{equation}}
\def\bea{\begin{eqnarray}}
\def\eea{\end{eqnarray}}
\def\bear{\begin{array}}
\def\ear{\end{array}}
\def\bfig{\begin{figure}}
\def\efig{\end{figure}}
\def\bcen{\begin{center}}
\def\ecen{\end{center}}
\def\bi{\begin{itemize}}
\def\ei{\end{itemize}}

\def\raw{\rightarrow}

\def\la{\label}

\begin{document}
\begin{titlepage}
\pubblock

\vfill
\Title{The physics of neutrino cross sections: theoretical studies}
\vfill
\Author{Luis Alvarez-Ruso}
\Address{Instituto de F\'\i sica Corpuscular (IFIC), Centro Mixto
  CSIC-UVEG, \\E-46071 Valencia, Spain}
\vfill
\begin{Abstract}
  The present status of neutrino cross section physics is reviewed focusing on the recent theoretical developments in quasielastic scattering, multi-nucleon contributions to the inclusive scattering and pion production on nucleons and nuclei. A good understanding of these processes is crucial to meet the precision needs of neutrino oscillation experiments. Some of the challenges that arise in the consistent description of MiniBooNE and MINERvA recent data are discussed. 
\end{Abstract}
\vfill
\begin{Presented}
 NuPhys2015: Prospects in Neutrino Physics

Barbican Centre, London, UK,  December 16--18, 2015

\end{Presented}
\vfill
\end{titlepage}
\def\thefootnote{\fnsymbol{footnote}}
\setcounter{footnote}{0}

\section{Introduction}

 Neutrino interactions offer unique opportunities for exploring 
 fundamental questions in astrophysics, nuclear and particle physics. Significant efforts are going into the study of the neutrino oscillation phenomenon, which has been established over the last 15 years. Neutrino oscillation experiments, currently evolving from the discovery to the precision stage, face a difficulty: the elusive nature of the neutrinos. Their presence and flavor can only be inferred by detecting the secondary particles created in the collisions with the nuclear targets used as detectors. A better understanding of neutrino interactions with nucleons and nuclei is crucial to distinguish signal from background and minimize systematic uncertainties in oscillation experiments.

\section{Quasielastic-like scattering}

Charged  current quasielastic (CCQE) scattering $\nu_l (\bar\nu_l) \, n(p) \raw  l^{\mp} \, p(n)$ is the most important neutrino interaction mechanism in the few-GeV energy region. It has been often used to identify the neutrino flavor and to determine its energy. Kinematic energy determination is based on the knowledge of the incoming neutrino direction, the measurement of outgoing lepton momentum, the assumption that the initial (bound) nucleon is at rest and the true CCQE nature of the event. On nuclear targets, the later relies on model dependent subtraction of irreducible CCQE-like backgrounds. This is important for the extraction of oscillation parameters because neutrino energy misreconstruction alters the position and depth of oscillation minima~\cite{Lalakulich:2012hs,Coloma:2013rqa,Ankowski:2016bji}.

CCQE scattering on nucleons is a conceptually well understood reaction. In the classic review of Ref.~\cite{LlewellynSmith:1971uhs}, the differential cross section (CS), in terms of the Mandelstam variables $s$, $t$ and $u$ is cast as
\be
\frac{d\sigma}{dt} = \frac{G_F^2 m_N^2}{8 \pi E_\nu^2} \left[ A(t) \pm B(t) \frac{(s-u)}{m_N^2}+ C(t) \frac{(s-u)^2}{m_N^4} \right]\,,
\ee
where $A$, $B$ and $C$ are quadratic functions of the vector [$F^V_{1,2}(t)$] and axial [$F_{A,P}(t)$] form factors (FF) (see for instance Eqs.~(58-60) of Ref.~\cite{Formaggio:2013kya}). Isospin symmetry allows to relate vector FF $F^V_{1,2}$ to the corresponding electromagnetic proton and neutron ones, which are extracted from electron scattering data. PCAC and the pion-pole dominance of the pseudoscalar FF, $F_P$, allow to express this FF in terms of $F_A$. In any case, the impact of  $F_P$ on the CCQE CS is minor except for $l=\tau$. The axial FF is usually parametrized as a dipole
\be
\la{eq:FA}
F_A (t) = g_A \left( 1 - \frac{t}{M_A^2} \right)^{-2} \,.
\ee
The value of $M_A$ extracted from early CCQE experiments on deuterium,  
$M_A = 1.016 \pm 0.026$~GeV~\cite{Bodek:2007ym}, is in agreement with the pion electroproduction result obtained from $\langle r_A^2 \rangle = 0.455 \pm 0.012$~fm$^2$~\cite{Bernard:1992ys} using that for the dipole ansatz $\langle r_A^2 \rangle = 12/M_A^2$. In spite of the fact that deviations from the dipole form have not been observed so far, it is worth stressing that the dipole  is not well justified from a theoretical point of view. Furthermore, it can be argued that the model dependent relation between $\langle r_A^2 \rangle$ and $M_A$ implies that $\langle r_A^2 \rangle$ is extracted from the whole experimentally available range of $t$ values, leading to an artificially small error. A new extraction of $F_A$ has been recently undertaken using a model-independent representation of the FF based on conformal mapping ($z$-expansion). The resulting  $\langle r_A^2 \rangle = 0.46(22)$~fm$^2$~\cite{Meyer:2016oeg} agrees with the value quoted above but with a much larger error. More precise determinations of $F_A(t)$ might become available from lattice QCD simulations in the future.

To model CCQE on nuclear targets, neutrino event generators have traditionally described the nucleus as a relativistic global Fermi gas (FG) of noninteracting nucleons. Such a simple picture fails in the quantitative description of electron scattering data. The variety of theoretical approaches developed over the last years include mean-field approximations~\cite{Maieron:2003df,Leitner:2006ww}, spectral functions~\cite{Benhar:2005dj,Ankowski:2007uy}, the Green function formalism~\cite{Meucci:2003cv}, RPA to deal with collective effects at low energy transfers~\cite{Singh:1992dc,Nieves:2004wx,Martini:2009uj} and continuum RPA that improves on the description of collective low-energy excitations~\cite{Pandey:2014tza}. The connection to electron-nucleus scattering has been phenomenologically explored by the superscaling approach~\cite{Caballero:2005sj}. It turns out that most of these more advanced descriptions from the nuclear-structure perspective predict integrated CCQE CS clearly smaller than the experimental ones reported by MiniBooNE~\cite{AguilarArevalo:2010zc} (see for instance Fig.~8 of Ref.~\cite{Alvarez-Ruso:2014bla}). There is now a consensus that a sizable contribution to the CS measured by MiniBooNE comes from amplitudes involving two interacting nucleons (2p2h excitations)~\cite{Nieves:2011yp,Martini:2011wp,Megias:2014qva}, as originally suggested in Ref.~\cite{Martini:2009uj}. As shown in Fig~\ref{fig:2diff}, the MiniBooNE flux averaged double-differential CS can be explained with a nucleon axial FF consistent with the measurements on deuterium ($M_A \sim 1$~GeV) once RPA and 2p2h terms are taken into account.
\bfig[h!]
\bcen
\includegraphics[width=0.41\linewidth]{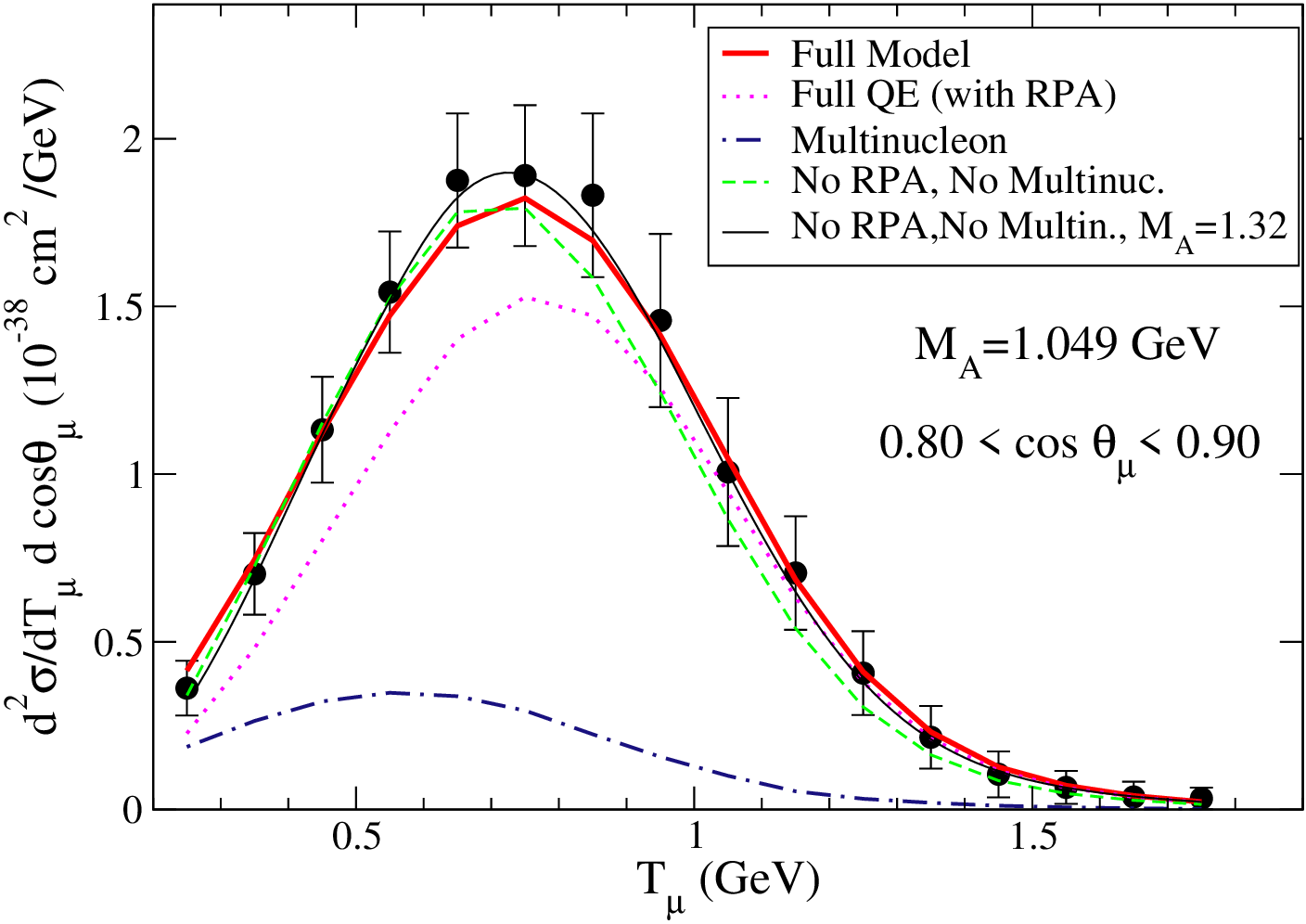}
\caption{Double differential cross section averaged over the MiniBooNE flux as predicted in Ref.~\cite{Nieves:2011yp}. The data of Ref.~\cite{AguilarArevalo:2010zc} have been rescaled by 0.9~\cite{Nieves:2011yp}, which is consistent with flux uncertainties. For a similar plot obtained with the model of Martini {\it et al.} see Fig.~6 of Ref.~\cite{Martini:2011wp}.} 
\label{fig:2diff}
\ecen
\efig 

Further evidence about the relevance of two-nucleon contributions has been provided by the {\it ab initio} Green's function Monte Carlo approach~\cite{Lovato:2014eva}. In these studies, the quantum many-body problem is solved for the full nuclear Hamiltonian with two- and three-body forces. Euclidean (imaginary time) responses are computed including one- and two-body currents. Sum rules contain a significant strength ($\sim 30$~\% in $^{12}$C) from two-nucleon terms~\cite{Lovato:2014eva}. The computational effort increases significantly with the number of nucleons, so the results are so far limited to light nuclei.     

In inclusive electron scattering, 2p2h mechanisms are clearly required to fill the dip region between the QE and $\Delta(1232)$ peaks~\cite{Gil:1997bm,Megias:2016lke,Rocco:2015cil} that are clearly visible as a function of the energy transfer for a low momentum transfer. Such a measurement with neutrinos has been undertaken by MINERvA~\cite{Rodrigues:2015hik}. As the neutrino energy is unknown, energy and momentum transfers cannot be determined solely from muon kinematics but the more challenging hadronic energy reconstruction is required. In Ref.~\cite{Rodrigues:2015hik}, data is compared to a GENIE implementation of the model of Ref.~\cite{Nieves:2011yp} with RPA correlations and 2p2h terms. This model improves the agreement with respect to the default one but some discrepancies remain. They could be due to deficiencies in the modeling of resonance excitation in nuclei and non-resonant pion production.

The availability of both MiniBooNE and MINERvA CCQE-like data opens the possibility to perform global fits with different models. This has been performed for the first time by the T2K Neutrino Interaction Working Group~\cite{Wilkinson:2016wmz}. The analysis reveals strong tensions between data sets. It was found that the NEUT implementation of the model of Ref.~\cite{Nieves:2011yp} describes the data best but, in order to fit the data, the 2p2h contribution had to be reduced by 27\% while $M_A$ was increased to 1.15~GeV. On the other hand, more flexible, consistent and kinematically complete implementations of theoretical models are required to render global fits more realistic and meaningful in the future.  

\section{Weak pion production}

Pion production is one of the main contributions to the inclusive CS in the energy range of interest for oscillation experiments. It can be part of the signal or a background that should be precisely constrained. Charge current interactions in which a primarily produced pion is absorbed are a source of QE-like events besides 2p2h. NC$1\pi^0$ events in Cherenkov detectors contribute to the $e$-like background in $\nu_e$ appearance measurements. Theoretically, $\pi$ production amplitudes enter the 2p2h models.

The first requirement for a precise description of neutrino induced pion production on nuclear targets is a realistic model at the nucleon level. Various studies have stressed the predominant role of the $\Delta(1232)3/2^+$ in the few-GeV region. The nucleon-to-$\Delta$ current can be written in terms of vector and axial transition FF, $C_{3-5}^V$ and $C_{3-6}^A$ in the notation of Ref.~\cite{LlewellynSmith:1971uhs}. Owing to the symmetry of the conserved vector current under isospin rotations, vector FF can be cast in terms of the helicity amplitudes extracted in the analysis of pion electroproduction data~\cite{Lalakulich:2005cs,Leitner:2008ue}. The information available about the axial part of the current is far more scarce. $C^A_5$ is the only FF that appears at leading order in an expansion of the hadronic tensor in the four-momentum transfer $q^2$. PCAC and the pion-pole dominance of the pseudoscalar FF $C_6^A$ allow to relate $C^A_5(0)$ to the  $\Delta N \pi$ effective coupling $g_{\Delta N\pi}$. In the chiral limit, $q_\mu \mathcal{A}^{\mu }_{N\Delta}=0$, resulting in the off-diagonal Goldberger-Treiman relation (GTR)
\be
C_5^A(0)\big{|}_{\mathrm{GTR}} =\sqrt{2/3}\, g_{\Delta N\pi} = 1.15 - 1.2  \,. 
\label{GT}
\ee
Deviations are expected only $\sim$ few \%, as they arise from chiral symmetry breaking. 

There have been several attempts to extract $C^A_5$ from ANL and BNL $\pi$ production bubble-chamber data on deuterium. In a model that  incorporated, besides the $\Delta(1232)$, non-resonant contributions at tree level complemented with phenomenological weak FF~\cite{Hernandez:2007qq}, $C_5^A(0) = 1.00\pm 0.11$ turned out to be 2$\sigma$ below the GTR value. Close to threshold, these non-resonant terms are fully determined by chiral symmetry. The model of Ref.~\cite{Hernandez:2010bx} could be reconciled with the GTR by simultaneously fitting vector FF to the electron-proton scattering structure function $F_2$~\cite{Graczyk:2014dpa}: $C_5^A(0) = 1.10^{+0.15}_{-0.14}$. However, it should be realized that the interplay between the real tree-level background and the complex $\Delta$ amplitudes violates Watson's theorem, which is a consequence of unitarity and time reversal invariance. By restoring Watson's theorem in the  most relevant $P_{33}$ partial wave it was found that 
\be
C_5^A(0) = 1.12 \pm 0.11 \,,
\label{fit1}
\ee
in agreement with the GTR of Eq.~(\ref{GT})~\cite{Alvarez-Ruso:2015eva}.

The determination of $C_5^A(0)$ suffers from long standing inconsistencies between the ANL and BNL data sets. A recent reanalysis~\cite{Wilkinson:2014yfa} has established that the origin of the discrepancies resides in the flux normalization. New consistent CS have then been obtained using flux-normalization independent CC1$\pi$/CCQE ratios. Taking advantage of these developments, a new fit has been performed in Ref.~\cite{Alvarez-Ruso:2015eva} leading to
\be
C_5^A(0) = 1.14\pm 0.07 \,.
\label{fit2}
\ee
This value is consistent with the one from the original data but in closer agreement with the GTR. Without imposing Watson's theorem, the revisited data lead to $C_5^A(0)=1.05 \pm 0.07$, clearly below GTR. This is in line with the findings of Ref.~\cite{Alam:2015gaa}, where a good description of the revised data is achieved with $C_5^A(0)=1$, using a model with heavier $N^*$ resonances but without unitarity. Remarkably, the consistency between the revised ANL and BNL data sets reduces the error in $C_5^A(0)$ from 10\% to 6\%. On the other hand, finer details in the structure of the pion production axial current are harder to pin down from these data. This is also the case for baryon resonances heavier than $\Delta(1232)$ that become more relevant as energy increases. Further insight requires new data on H$_2$/D$_2$ targets or indirectly on multinuclear targets~\cite{Lu:2015hea}.

A state of the art description of meson production by a dynamical model in coupled channels has been recently extended to weak reactions~\cite{Nakamura:2015rta}. PCAC is used to derive the axial current. The full amplitudes are the solution of the Lippmann-Schwinger equation. For these reasons, GTR and the Watson's theorem are respected by construction. The total CS in the $\nu_\mu p\to \mu^- p \pi^+$ channel is higher than the reanalyzed data of Ref.~\cite{Wilkinson:2014yfa} but deuteron corrections were not considered in Ref.~\cite{Nakamura:2015rta}. Within the spectator approximation, deuteron effects cause a small reduction ($< 8$\%)~\cite{AlvarezRuso:1998hi}, that would make the agreement better. Ref.~\cite{Wu:2014rga} obtains an additional reduction at forward angles from the strong interaction of outgoing $p n$ pairs. This calls for a more detailed analysis, accounting for ANL and BNL kinematical cuts. 

Modeling pion production on nuclei carry additional challenges. The initial nucleon is often assumed to be free, with a Fermi momentum chosen according to the global or local FG models. More elaborated descriptions of the initial state like spectral functions~\cite{Benhar:2005dj} and  bound-sate wave functions~\cite{Praet:2008yn} have also become available for baryon resonance excitation and meson production. Nevertheless at the higher energy transfers of inelastic processes, the details of nuclear structure should be less relevant. The hadronic currents are also modified in the nucleus. The main effect is the increase of the $\Delta(1232)$  width (broadening) by many-body processes: $\Delta \,N \raw N \, N$, $\Delta \,N \raw N \, N \, \pi$, $\Delta \,N \, N \raw N \, N \, N$. In their way out of the nucleus, pions also undergo final state interactions (FSI): they can be absorbed, change their energy, angle and charge. In CC interactions, there is a considerable side feeding from the dominant $\pi^+$ production to the $\pi^0$ channel~\cite{Leitner:2006ww}. At high momentum transfers, low energy pions can also be produced in secondary collisions of nucleons knocked out in QE interactions~\cite{Leitner:2006ww}. The imprint of the strong-interacting environment on the observables is therefore quite significant, obscuring the connection between primary interactions and measured quantities.  

\begin{figure}[h!]
\begin{center}
\vspace{2mm}    
\includegraphics[width=0.4\textwidth]{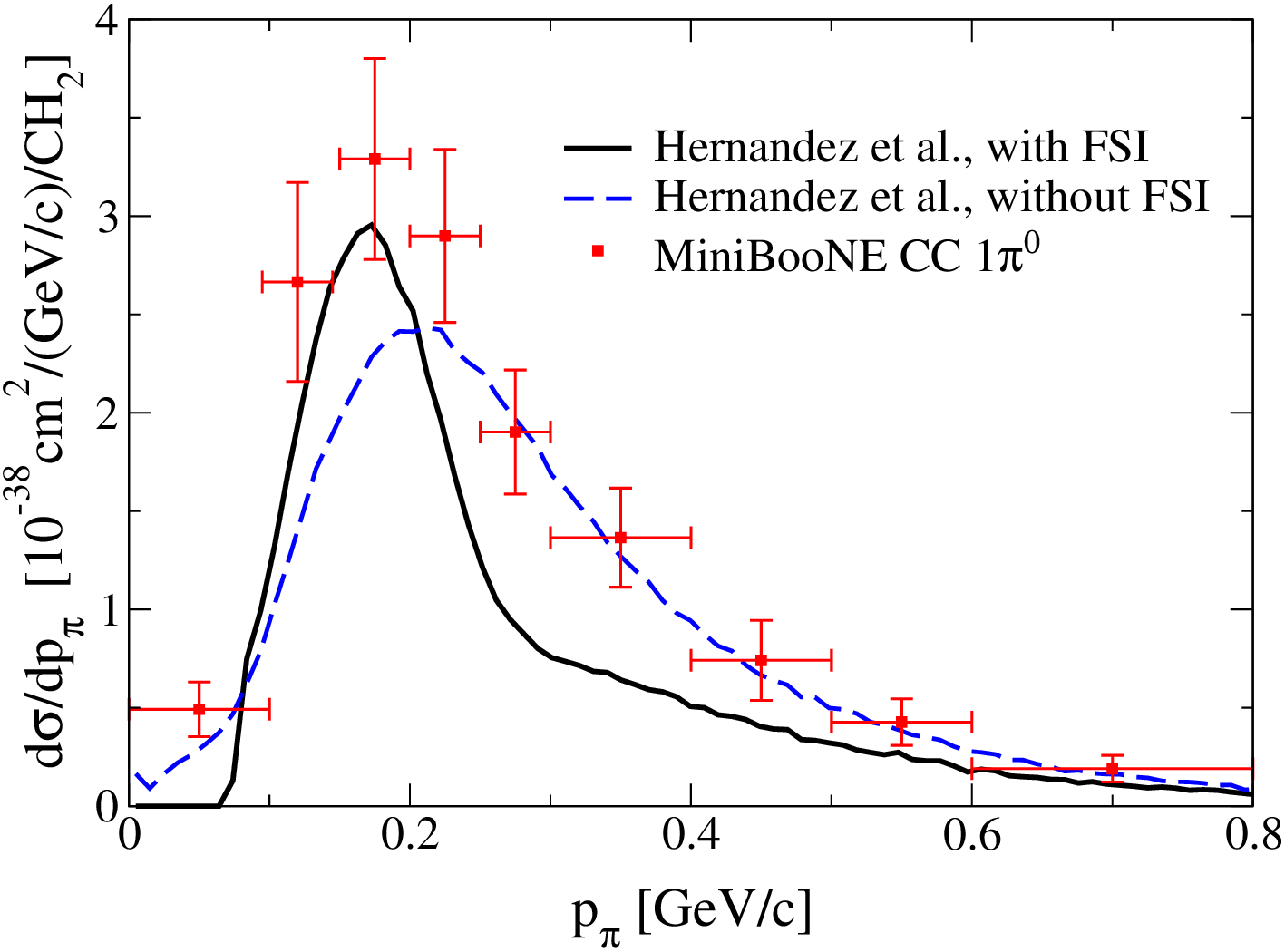}\hspace{0.05\textwidth}\vspace{-4mm} \includegraphics[width=0.41\textwidth]{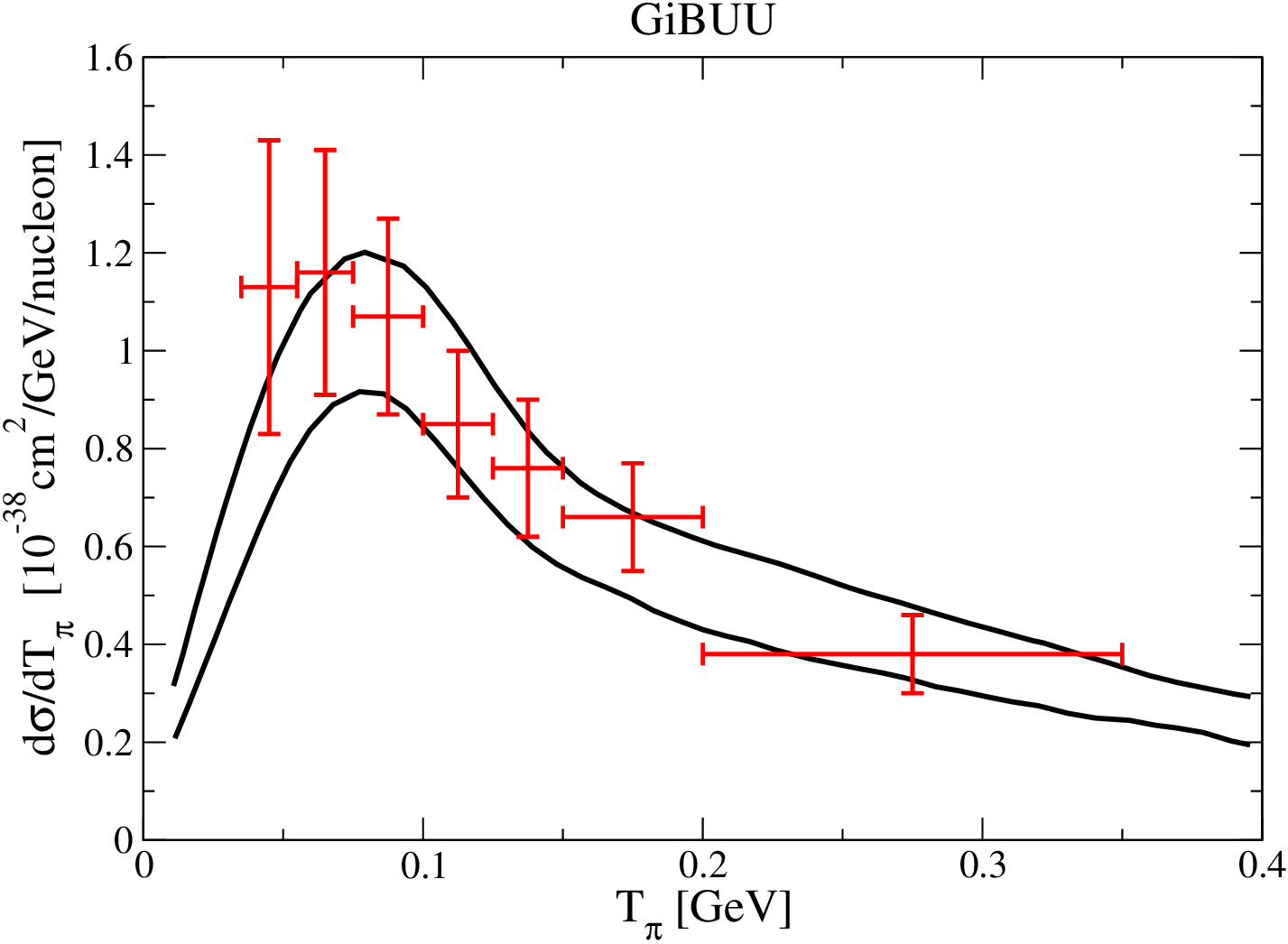}
\end{center}
\caption{Left: CC1$\pi^0$ on CH$_2$ folded with the $\nu_\mu$ flux at the MiniBooNE detector. Data from Ref.~\cite{AguilarArevalo:2010xt} are compared to the cascade approach of Ref.~\cite{Hernandez:2013jka}. Right (adapted from Ref.~\cite{Mosel:2015tja}): Differential CS for CC$\pi^\pm$ on CH averaged over the MINERvA flux, computed with the GiBUU transport model. Experimental results are from Ref.~\cite{Eberly:2014mra}.}
\label{fig:CCpi}
\end{figure}

The MiniBooNE measurements, reported as single pion momentum and angular flux-averaged distributions, have been compared to the most comprehensive approaches available~\cite{Lalakulich:2012cj,Hernandez:2013jka}. In spite of the different treatment of FSI, the results are very similar. The comparison of the model of Ref.~\cite{Hernandez:2013jka}, for example, to data, displayed in Fig.~\ref{fig:CCpi} (left) for CC1$\pi^0$, reveals an unexplained excess of pions. Such shrinking of the peak by FSI has however been observed in pion photoproduction~\cite{Krusche:2004uw}.
The shape disagreement apparent in Fig~\ref{fig:CCpi} (left) is in contrast with the result of the GiBUU model for CC$\pi^{\pm}$ (mostly $\pi^+$) reaction compared to MINERvA data,  Fig~\ref{fig:CCpi} (right). One is tempted to attribute the different scenarios  in Fig~\ref{fig:CCpi} to the $\nu$ fluxes: the flux at MiniBooNE peaks at around 700~MeV while the MINERvA one does close to 3~GeV. However, according to Ref.~\cite{Sobczyk:2014xza}, there should be a strong correlation among the two data sets in spite of the flux differences. Using the NuWro generator, the authors of Ref.~\cite{Sobczyk:2014xza} have obtained that the ratio $\left(d\sigma / dT_\pi\right)_{\mathrm{MINERvA,\,CC}\pi^\pm} / {\left(d\sigma / dT_\pi\right)_{\mathrm{MiniBooNE,\,CC}\pi^+}}$ is approximately constant but such a correlation is absent in the data (see Fig.~6 of Ref.~\cite{Sobczyk:2014xza}). Further progress in the understanding of weak pion production requires this tension to be resolved.

\Acknowledgements
I wish to thank the NuPhys2015 organizers for the opportunity to present this overview. I acknowledge financial support by the Spanish Ministerio de Econom\'\i a y Competitividad and the European Regional Development Fund, under contracts  FIS2014-51948-C2-1-P and SEV-2014-0398, and by  Generalitat Valenciana under contract PROMETEOII/2014/0068.

\end{document}